\documentclass[prl,twocolumn,showpacs,10pt]{revtex4-1}
\usepackage{amsmath}
\usepackage{amsfonts}
\usepackage{amssymb}
\usepackage{mathrsfs}
\usepackage{latexsym}
\usepackage{dsfont}
\usepackage{graphicx}

\newcommand{\tr}{{\rm Tr}}

\newenvironment{proof}[1][Proof]{\noindent\textbf{#1.} }{\ \rule{0.5em}{0.5em}}

\begin{document}

\title{witness to detect the genuine multipartite semiquantum for arbitrary N partite quantum state. }
\author{Zhi-Hao Ma}
\email[Email:]{ma9452316@gmail.com}
\affiliation{Department of Mathematics, Shanghai Jiaotong
University, Shanghai, 200240, P.R.China}

\author{Zhi-Hua Chen}
\affiliation{Department of Science, Zhijiang college, Zhejiang
University of technology, Hangzhou, 310024, P.R.China}

\date{\today }

\begin{abstract}
witness to detect the genuine multipartite semiquantum for arbitrary N partite quantum state.

\end{abstract}


\maketitle

%%%%%%%%%%%%%%%%%%%%%%%%%%%%%%%%%%%%%%%%%%%%%%%%%%%%%%%%%%%%%%%%%%%%%%%%%%

Entanglement, is a distinctive feature of quantum mechanics \cite{Horodecki09,Guhne09}, and has been found numerous applications in quantum information processing tasks
\cite{Nielsen:00}.The problem of detect whether a quantum state is entanglement or not is widely studied,  However, entanglement does not necessarily exhaust all quantum correlations present in a state. Beyond entanglement, quantum discord is a suitable measure of quantum correlation.The total correlations between two quantum systems $A$ and $B$ are quantified by the quantum mutual information
\begin{equation}\label{Totale}
    \mathcal{I}(\rho_{AB})=S(\rho_{A})+S(\rho_{B})-S(\rho_{AB})
\end{equation}
where $S(\rho)=-\mathrm{Tr}(\rho \log_2 \rho)$ is the Von Neumann entropy and $\rho_{A(B)}= \mathrm{Tr}_{B(A)}(\rho_{AB})$.

On the other hand, the classical part of correlations is defined as the maximum information about one subsystem that can be obtained by performing a measurement on the other system. Given a set of projective (von Neumann) measurements described by a complete set of orthogonal projectors $\{\hat{\Pi}_{B}^{j}\}=\{|b_j\rangle\langle b_j|\}$ and locally performed only on system $B$, which satisfying that $\hat{\Pi}_{B}^{j}\geqslant0$, $\sum_k \hat{\Pi}_{B}^{j}=I$, $I$ is the identity operator, then the information about $A$ is the difference between the initial entropy of $A$ and the conditional entropy, that is $\mathcal{I}(\rho_{AB}|\{\hat{\Pi}_{B}^{j}\})=S(\rho_{A})-\sum_j p_j S(\rho_j)$, where $\rho_j=(I\otimes \hat{\Pi}_{B}^{j})\rho(I\otimes \hat{\Pi}_{B}^{j})/\mathrm{Tr}[(I\otimes\hat{\Pi}_{B}^{j})\rho(I\otimes\hat{\Pi}_{B}^{j})]$, $p_j$ is the probability of the measurement outcome $j$ and $I$ is the identity operator for subsystem $A$. Classical correlations are thus quantified by $\mathcal{Q}(\rho_{AB})=\mathrm{sup}_{\{\hat{\Pi}_{B}^{j}\}}\mathcal{I}(\rho_{AB}|\{\hat{\Pi}_{B}^{j}\})$ and the quantum discord is then defined by
\begin{equation}\label{quantum discord}
    \mathcal{D_{B}}(\rho_{AB})=\mathcal{I}(\rho_{AB})- \mathcal{Q}(\rho_{AB}),
\end{equation}
which is zero only for states with classical correlations and nonzero for states with quantum correlations. The nonclassical correlations captured by the quantum discord may be present even in separable states.\cite{Ollivier:01}

{\it Relative entropy and symmetric quantum discord. --}
The quantum relative entropy is a measure of distinguishability
between two arbitrary density operators $\hat{\rho}$ and $\hat{\sigma}$, which is defined as
$S\left( \hat{\rho}\parallel \hat{\sigma}\right) =\mathrm{Tr}\left( \hat{\rho}%
\log_2 \hat{\rho}-\hat{\rho}\log_2 \hat{\sigma}\right)$~\cite{Vedral:02}.
We can express the quantum mutual information $I(\hat{\rho}_{AB})$
as the relative entropy between $\hat{\rho}_{AB}$ and the product state
$\hat{\rho}_{A}\otimes\hat{\rho}_{B}$, i.e.
\begin{equation}
I\left( \hat{\rho}_{AB}\right) =S\left( \hat{\rho}_{AB}\parallel \hat{\rho}_{A}\otimes
\hat{\rho}_{B}\right).
\end{equation}
In order to express the measurement-induced quantum mutual information
$J\left( \hat{\rho}_{AB}\right)$ in terms of relative entropy, we need to consider
a non-selective von Neumann measurement on part $B$ of $\hat{\rho}_{AB}$,
which yields
%%v2
$\Phi _{B}\left( \hat{\rho}_{AB}\right) = \sum_{j}
\left( \hat{1}_{A}\otimes \hat{\Pi}_{B}^{j} \right)
\hat{\rho}_{AB}
\left(\hat{1}_{A}\otimes \hat{\Pi}_{B}^{j}\right)
=\sum_{j}p_{j}\hat{\rho}_{A|j}\otimes \left\vert b_{j}\right\rangle
\left\langle b_{j}\right\vert$.
Moreover, tracing over the variables of the subsystem $A$, we obtain
%%v2
$\Phi _{B}\left( \hat{\rho}_{B}\right) =
\Phi_{B}\left( \mathrm{Tr}_{A}\,\hat{\rho}_{AB}\right)
=\sum_{j}p_{j}\left\vert b_{j}\right\rangle \left\langle b_{j}\right\vert$,
where we have used that $\mathrm{Tr}_{A} (\hat{\rho}_{A|j})=1$. Then,
by expressing the entropies
$S\left( \Phi _{B}\left( \hat{\rho}_{AB}\right)\right) $ and
$S\left( \Phi _{B}\left( \hat{\rho}_{B}\right) \right) $ as
$S\left( \Phi _{B}\left( \hat{\rho}_{AB}\right) \right) = H\left( \mathbf{p}%
\right) +\sum_{j}p_{j}S\left( \hat{\rho}_{A|j}\right)$ and
$S\left( \Phi _{B}\left( \hat{\rho}_{B}\right) \right) =H\left( \mathbf{p}%
\right)$, with $H\left( \mathbf{p}\right)$ denoting the Shannon entropy
$H\left( \mathbf{p}\right) =-\sum_{j}p_{j}\log _{2}\left( p_{j}\right)$,
we can rewrite $J(\hat{\rho}_{AB})$ as
\begin{equation}
J\left( \hat{\rho}_{AB}\right) = S\left( \Phi _{B}\left( \hat{\rho}_{AB}\right) \parallel \hat{\rho}%
_{A}\otimes \Phi _{B}\left( \hat{\rho}_{B}\right) \right).
\end{equation}
Therefore, the quantum discord can be rewriten in terms of a
difference of relative entropies:
$\overline{\mathcal{D}}\left( \hat{\rho}_{AB}\right) = S\left( \hat{\rho}_{AB}\parallel \hat{\rho}%
_{A}\otimes \hat{\rho}_{B}\right) -S\left( \Phi _{B}\left( \hat{\rho}%
_{AB}\right) \parallel \hat{\rho}_{A}\otimes \Phi _{B}\left( \hat{\rho}%
_{B}\right) \right)$,
%%v2
with minimization taken over $\{\hat{\Pi}_{B}^{j}\}$ to remove
the measurement-basis dependence.
It is possible then to obtain a natural symmetric extension $\mathcal{D}\left(\hat{\rho}_{AB}\right)$
for the quantum discord $\overline{\mathcal{D}}\left( \hat{\rho}_{AB}\right)$.

Indeed, performing measurements over  subsystem
$A$, we define $A$-discord as:

\begin{eqnarray}
\mathcal{D_{A}}\left( \hat{\rho}_{AB}\right) &=& \min_{\{\hat{\Pi}_{A}^{k}\otimes I\}}
\left[ S\left( \hat{\rho}_{AB}\parallel \Phi
^{A}_{AB}\left( \hat{\rho}_{AB}\right) \right) \right. \nonumber \\
&&\left.\hspace{-0.6cm}-S\left( \hat{\rho}_{A}\parallel
\Phi_{A}\left( \hat{\rho}_{A}\right) \right) -S\left( \hat{\rho}%
_{B}\parallel \Phi^{I}_{B}\left( \hat{\rho}_{B}\right) \right)\right] .
\end{eqnarray}
where the operator $\Phi^{A}_{AB}$ is given by
\begin{equation}
\Phi^{A}_{AB}\left( \hat{\rho}_{AB}\right) =\sum_{k} \left(\hat{\Pi}_{A}^{k}\otimes I
 \right) \hat{\rho}_{AB} \left(\hat{\Pi}_{A}^{k}\otimes I\right) \, .
\end{equation}

and the operator $\Phi_{A}$ is given by
\begin{equation}
\Phi_{A}\left( \hat{\rho}_{A}\right) =\sum_{k} \left(
\hat{\Pi}_{A}^{k} \right) \hat{\rho}_{A} \left( \hat{\Pi}_{A}^{k}\right) \, .
\end{equation}

and the operator $\Phi^{I}_{B}$ is in fact the identity operator
\begin{equation}
\Phi^{I}_{B}\left( \hat{\rho}_{B}\right) =\hat{\rho}_{B}
\end{equation}

performing measurements over  subsystem
$B$, we define $B$-discord as:
\begin{eqnarray}
\mathcal{D_{B}}\left( \hat{\rho}_{AB}\right) &=& \min_{\{I\otimes\hat{\Pi}_{B}^{k}\}}
\left[ S\left( \hat{\rho}_{AB}\parallel \Phi
^{B}_{AB}\left( \hat{\rho}_{AB}\right) \right) \right. \nonumber \\
&&\left.\hspace{-0.6cm}-S\left( \hat{\rho}_{A}\parallel
\Phi^{I}_{A}\left( \hat{\rho}_{A}\right) \right) -S\left( \hat{\rho}%
_{B}\parallel \Phi _{B}\left( \hat{\rho}_{B}\right) \right)\right] .
\end{eqnarray}
where the operator $\Phi^{B}_{AB}$ is given by
\begin{equation}
\Phi^{B}_{AB}\left( \hat{\rho}_{AB}\right) =\sum_{k} \left(I\otimes
\hat{\Pi}_{B}^{k} \right) \hat{\rho}_{AB} \left(I\otimes \hat{\Pi}_{B}^{k}\right) \, .
\end{equation}

and the operator $\Phi_{B}$ is given by
\begin{equation}
\Phi_{B}\left( \hat{\rho}_{B}\right) =\sum_{k} \left(
\hat{\Pi}_{B}^{k} \right) \hat{\rho}_{B} \left( \hat{\Pi}_{B}^{k}\right) \, .
\end{equation}

and the operator $\Phi^{I}_{A}$ is in fact the identity operator
\begin{equation}
\Phi^{I}_{A}\left( \hat{\rho}_{A}\right) =\hat{\rho}_{A}
\end{equation}

Then, the symmetric discord is defined as \begin{equation}\mathcal{D}(\hat{\rho}_{AB})=\min[\mathcal{D}_{A}(\hat{\rho}_{AB}),\mathcal{D}_{B}(\hat{\rho}_{AB})]
\label{DD}\end{equation}

The aim of this work is to give a measure of genuine multipartite quantum discord for arbitrary N partite state.
We will extend quantum discord as given by Eq.~(\ref{DD}) to multipartite systems.

Recall that an $N$-partite pure state $|\psi\rangle\in \mathcal{H}_1\otimes \mathcal{H}_2\otimes\cdots
\mathcal{H}_N$ is called biseparable if there is a bipartition
$j_1j_2\cdots j_k|j_{k+1}\cdots j_N$ such that
\begin{equation}\label{} |\psi\rangle=|\psi_1\rangle_{j_1j_2\cdots
j_k}|\psi_2\rangle_{j_{k+1}\cdots j_N},
\end{equation}
where $\{j_1,j_2,\cdots j_{k}|j_{k+1},\cdots j_N \}$ is any
partition of $\{1,2,\cdots, N\}$, e.g., $\{13|24\}$ is a partition
of $\{1,2,3,4\}$.

Let $\gamma$ be any subset $\{j_{1}j_{2}\cdots j_{k}\}$ of $\{1, 2,..., N\}$, corresponding to
a partition $j_{1}j_{2}\cdots j_{k}|j_{k+1}\cdots j_{N}$, e.g., for three qubits state,
$\gamma=1$ corresponding to the partition $A|BC$, and corresponding to the reduced density matrix $\rho_{A}$, while if $\gamma=23$, then it corresponding to the reduced density matrix $\rho_{BC}$.

{\bf Definition 1.} For an arbitrary $N$ partite state $\hat{\rho}_{1 \cdots N}$, the genuine multipartite quantum discord $\mathcal{D}\left( \hat{\rho}_{1 \cdots N} \right)$ is defined as follows:

(1). First, let $\rho$ be an n partite  state, and $\gamma$ be any subset $\{j_{1}j_{2}\cdots j_{k}\}$ of $\{1, 2,..., N\}$, corresponding to
a partition $j_{1}j_{2}\cdots j_{k}|j_{k+1}\cdots j_{N}$, e.g., for three qubits state,
$\gamma=1$ corresponding to the partition $A|BC$, and corresponding to the reduced density matrix $\rho_{A}$, and $\gamma^{\prime}$ is defined as the complemental set of $\gamma$(that is, the set union of $\gamma$ and $\gamma^{\prime}$ is the total set $\{1, 2,..., N\}$, i.e.,for three qubits state, if $\gamma=1$, then $\gamma^{\prime}=23$), then define the $\gamma$-discord as

\begin{eqnarray}
\mathcal{D_{\gamma}}\left( \hat{\rho}_{1, 2,..., N}\right) &=& \min_{\{I_{\gamma^{\prime}}\otimes\hat{\Pi}_{\gamma}^{k}\}}
\left[ S\left( \hat{\rho}_{1, 2,..., N}\parallel \Phi
^{\gamma}_{1, 2,..., N}\left( \hat{\rho}_{1, 2,..., N}\right) \right) \right. \nonumber \\
&&\left.\hspace{-0.6cm} -S\left( \hat{\rho}%
_{\gamma}\parallel \Phi _{\gamma}\left( \hat{\rho}_{\gamma}\right) \right)\right] .
\end{eqnarray}
where the operator $\Phi^{\gamma}_{1, 2,..., N}$ is given by
\begin{equation}
\Phi^{\gamma}_{1, 2,..., N}\left( \hat{\rho}_{1, 2,..., N}\right) =\sum_{k} \left(I_{\gamma^{\prime}}\otimes
\hat{\Pi}_{\gamma}^{k} \right) \hat{\rho}_{1, 2,..., N} \left(I_{\gamma^{\prime}}\otimes \hat{\Pi}_{\gamma}^{k}\right) \, .\label{super}
\end{equation}

Here $I_{\gamma^{\prime}}$ is the identity operator for subsystem $\gamma^{\prime}$,
and the operator $\Phi_{\gamma}$ is given by
\begin{equation}
\Phi_{\gamma}\left( \hat{\rho}_{\gamma}\right) =\sum_{k} \left(
\hat{\Pi}_{\gamma}^{k} \right) \hat{\rho}_{\gamma} \left( \hat{\Pi}_{\gamma}^{k}\right) \, .
\end{equation}

(2). then define the genuine multipartite quantum discord  as the minimal of all $\gamma$-discord:
\begin{equation}
\mathcal{D}\left( \hat{\rho}_{1, 2,..., N}\right) =\min_{\gamma}\mathcal{D_{\gamma}}\left( \hat{\rho}_{1, 2,..., N}\right)
\end{equation}

where the min run over all partition $\gamma$.

Take three partite quantum state $\rho_{123}$ as example.

For $\gamma=1$,

\begin{eqnarray}
\mathcal{D}_{1}\left( \hat{\rho}_{1, 2,3}\right) &=& \min_{\{I_{23}\otimes\hat{\Pi}_{1}^{k}\}}
\left[ S\left( \hat{\rho}_{1, 2,3}\parallel \Phi
^{1}_{1, 2,3}\left( \hat{\rho}_{1, 2,3}\right) \right) \right. \nonumber \\
&&\left.\hspace{-0.6cm} -S\left( \hat{\rho}%
_{1}\parallel \Phi _{1}\left( \hat{\rho}_{1}\right) \right)\right] .
\end{eqnarray}
where the operator $\Phi^{1}_{1, 2,3}$ is given by
\begin{equation}
\Phi^{1}_{1, 2,3}\left( \hat{\rho}_{1, 2,3}\right) =\sum_{k} \left(\hat{\Pi}_{1}^{k}\otimes I_{23}
 \right) \hat{\rho}_{1, 2,3} \left(\hat{\Pi}_{1}^{k}\otimes I_{23}\right) \, .
\end{equation}

and the operator $\Phi_{1}$ is given by
\begin{equation}
\Phi_{1}\left( \hat{\rho}_{1}\right) =\sum_{k} \left(
\hat{\Pi}_{1}^{k} \right) \hat{\rho}_{1} \left( \hat{\Pi}_{1}^{k}\right) \, .
\end{equation}

For $\gamma=2$,

\begin{eqnarray}
\mathcal{D}_{2}\left( \hat{\rho}_{1, 2,3}\right) &=& \min_{\{I_{13}\otimes\hat{\Pi}_{2}^{k}\}}
\left[ S\left( \hat{\rho}_{1, 2,3}\parallel \Phi
^{2}_{1, 2,3}\left( \hat{\rho}_{1, 2,3}\right) \right) \right. \nonumber \\
&&\left.\hspace{-0.6cm} -S\left( \hat{\rho}%
_{2}\parallel \Phi _{2}\left( \hat{\rho}_{2}\right) \right)\right] .
\end{eqnarray}
where the operator $\Phi^{2}_{1, 2,3}$ is given by
\begin{equation}
\Phi^{2}_{1, 2,3}\left( \hat{\rho}_{1, 2,3}\right) =\sum_{k} \left(I_{1}\otimes \hat{\Pi}_{2}^{k}\otimes I_{3}
 \right) \hat{\rho}_{1, 2,3} \left(I_{1}\otimes \hat{\Pi}_{2}^{k}\otimes I_{3}\right) \, .
\end{equation}

and the operator $\Phi_{2}$ is given by
\begin{equation}
\Phi_{2}\left( \hat{\rho}_{2}\right) =\sum_{k} \left(
\hat{\Pi}_{2}^{k} \right) \hat{\rho}_{2} \left( \hat{\Pi}_{2}^{k}\right) \, .
\end{equation}

For $\gamma=12$,

\begin{eqnarray}
\mathcal{D}_{12}\left( \hat{\rho}_{1, 2,3}\right) &=& \min_{\{I_{3}\otimes\hat{\Pi}_{12}^{k}\}}
\left[ S\left( \hat{\rho}_{1, 2,3}\parallel \Phi
^{12}_{1, 2,3}\left( \hat{\rho}_{1, 2,3}\right) \right) \right. \nonumber \\
&&\left.\hspace{-0.6cm} -S\left( \hat{\rho}%
_{12}\parallel \Phi _{12}\left( \hat{\rho}_{12}\right) \right)\right] .
\end{eqnarray}
where the operator $\Phi^{12}_{1, 2,3}$ is given by
\begin{equation}
\Phi^{12}_{1, 2,3}\left( \hat{\rho}_{1, 2,3}\right) =\sum_{k} \left( \hat{\Pi}_{12}^{k}\otimes I_{3}
 \right) \hat{\rho}_{1, 2,3} \left( \hat{\Pi}_{12}^{k}\otimes I_{3}\right) \, .
\end{equation}

and the operator $\Phi_{12}$ is given by
\begin{equation}
\Phi_{12}\left( \hat{\rho}_{12}\right) =\sum_{k} \left(
\hat{\Pi}_{12}^{k} \right) \hat{\rho}_{12} \left( \hat{\Pi}_{12}^{k}\right) \, .
\end{equation}

Similarly, we can get the case of $\gamma=3,\gamma=13,\gamma=23$, and get $\mathcal{D}\left( \hat{\rho}_{1, 2,3}\right)=\min_{\gamma}\mathcal{D_{\gamma}}\left( \hat{\rho}_{1, 2,3}\right)$, where $\gamma$ run over the six sets $1,2,3,12,13,23$.

{\bf Theorem 1.} For an $N$ partite quantum state  $\hat{\rho}_{A_{1} \cdots A_{N}}$  on Hilbert space $H_{1}\otimes H_{2}\cdots H_{N}$, The genuine multipartite quantum discord $\mathcal{D}\left( \hat{\rho}_{A_{1} \cdots A_{N}} \right)$ is
non-negative, i.e., $\mathcal{D}\left( \hat{\rho}_{A_{1} \cdots A_{N}}  \right) \geqslant 0$.

\begin{proof} To prove that $\mathcal{D}\left( \hat{\rho}_{A_1 \cdots A_N}  \right) \geqslant 0$,we need to prove that, for any $\gamma$, the $\gamma$-discord is non-negative, i.e., $\mathcal{D_{\gamma}}\left( \hat{\rho}_{A_1 \cdots A_N}  \right) \geqslant 0$. Without loss of generality, we can assume that $\gamma=1$, which corresponding to the reduced density matrix $\hat{\rho}_{A_{1}}$.
Define the POVM as follows: define
$\Phi\left( \hat{\rho}_{A_1 \cdots A_N} \right) = \sum_{k} {\hat{\Pi}}_{k} \, \hat{\rho}_{A_1 \cdots A_N} \,
{\hat{\Pi}}_{k}$,
with $\hat{\Pi}_{k} = \hat{\Pi}_{A_1}^{j_1} \otimes \cdots \otimes \hat{\Pi}_{A_N}^{j_N}$ and
$k$ denoting the index string $(j_1 \cdots j_N$), with $j_1$ an arbitrary integer, $j_2=j_3=\cdots j_N=1$, and define $\Phi_1\left( \hat{\rho}_{A_1} \right) = \sum_{j_1} \hat{\Pi}_{A_1}^{j_1} \, \hat{\rho}_{A_1} \,
\hat{\Pi}_{A_1}^{j_1}$ and $\Phi_i\left( \hat{\rho}_{A_i} \right) = \hat{\rho}_{A_i}, i=2,3\cdots N$.

We associate with each subsystem $A_j$ an ancilla system $B_j$. Therefore, we will define
a composite density operator $\hat{\rho}_{A_1 \cdots A_N;B_1\cdots B_N}^\prime$ such that
\begin{equation}
\hat{\rho}_{A_1 \cdots A_N;B_1\cdots B_N}^\prime = \sum_k \sum_{k^\prime}
{\hat{\Pi}}_{k} \, \hat{\rho}_{A_1 \cdots A_N} \, {\hat{\Pi}}_{k^\prime} \otimes
\hat{\Lambda}_{k k^\prime},
\label{rho-ancilla}
\end{equation}
where $\hat{\Lambda}_{k k^\prime} = |B_{j_1}\cdots B_{j_N}\rangle \langle B_{j^\prime_1}\cdots B_{j^\prime_N}|$,
with $k$ and $k^\prime$ denoting the index strings $(j_1 \cdots j_N$) and $(j^\prime_1 \cdots j^\prime_N$),with $j_1,j^\prime_1$  arbitrary integers, $j_2=j_3=\cdots j_N=j^\prime_1=j^\prime_2=\cdots=j^\prime_N=1$,
respectively. From the monotonicity of the relative entropy under partial trace~\cite{Ruskai:02},
for any positive operators $\hat{\sigma}_{12}$ and $\hat{\gamma}_{12}$ such that
$\mathrm{Tr}\left( \hat{\sigma}_{12}\right) =\mathrm{Tr}\left( \hat{\gamma}_{12}\right)$,
we have that $S\left( \hat{\sigma}_{12}\Vert \hat{\gamma}_{12}\right) \geqslant S\left(
\hat{\sigma}_{1}\Vert \hat{\gamma}_{1}\right)$, where
$\hat{\sigma}_{1} =\mathrm{Tr}_{2}\left( \hat{\sigma}_{12}\right)$ and
$\hat{\gamma}_{1} =\mathrm{Tr}_{2}\left( \hat{\gamma}_{12}\right)$. Then
$S\left( \hat{\sigma}_{123\ldots N}\Vert \hat{\gamma}_{123\ldots N}\right)
\geqslant \ldots \geqslant S\left( \hat{\sigma}_{123}\Vert \hat{\gamma}_{123}\right)
\geqslant S\left( \hat{\sigma}_{12}\Vert \hat{\gamma}_{12}\right) \geqslant S\left( \hat{\sigma}_{1}\Vert \hat{\gamma}_{1}\right)$.
By taking $\hat{\rho}_{A_1 \cdots A_N;B_1\cdots B_N}^\prime$ as $\hat{\sigma}$ and
$\hat{\rho}_{A_1;B_1}\otimes \hat{\rho}_{A_2;B_2}^\prime \otimes \ldots \otimes \hat{\rho}_{A_N;B_N}^\prime$
as $\hat{\gamma}$, we obtain
\begin{eqnarray}
&&\hspace{-1.3cm}S\left( \hat{\rho}_{A_1 \cdots A_N;B_1\cdots B_N}^\prime \Vert
\hat{\rho}_{A_1;B_1}^\prime \otimes \hat{\rho}_{A_2;B_2}^\prime \otimes \ldots \otimes \hat{\rho}_{A_N;B_N}^\prime \right) \nonumber \\
\hspace{1cm} &\geqslant& S\left( \hat{\rho}_{A_1 \cdots A_N}^\prime \Vert \hat{\rho}_{A_1}^\prime \otimes \ldots \otimes \hat{\rho}_{A_N}^\prime \right),
\label{ineq-1}
\end{eqnarray}
which therefore implies that
$\sum_{j=1}^N S\left( \hat{\rho}_{A_j;B_j}^\prime \right) -
S\left( \hat{\rho}_{A_1 \cdots A_N;B_1\cdots B_N}^\prime \right)
\geqslant \sum_{j=1}^N S\left( \hat{\rho}_{A_j}^\prime \right) -
S\left( \hat{\rho}_{A_1 \cdots A_N}^\prime \right)$.
Moreover, from Eq.~(\ref{rho-ancilla}), it follows the relations:
$S\left( \hat{\rho}_{A_1 \cdots A_N;B_1\cdots B_N}^\prime \right) = S\left( \hat{\rho}_{A_1 \cdots A_N}\right)$,
$S\left( \hat{\rho}_{A_1 \cdots A_N}^{\prime }\right) = S\left( \Phi \left( \hat{\rho}_{A_1 \cdots A_N}\right) \right)$,
$S\left( \hat{\rho}_{A_j;B_j}^\prime \right) = S\left( \hat{\rho}_{A_j} \right) \,\, (\forall j)$, and
$S\left( \hat{\rho}_{A_j}^{\prime }\right) = S\left( \Phi _{j}\left( \hat{\rho}_{A_j}\right) \right) \,\, (\forall j)$.
Then, inequality~(\ref{ineq-1}) becomes
$\hspace{-0.3cm}\sum_{j=1}^{N} S\left( \hat{\rho}_{A_j}\right) - S\left( \hat{\rho}_{A_1 \cdots A_N} \right)
\geqslant \sum_{j=1}^{N} S\left( \Phi_{j}\left( \hat{\rho}_{A_j }\right) \right)
- S\left( \Phi \left( \hat{\rho}_{A_1 \cdots A_N}\right)  \right)$.
By rewriting this inequality in terms of the relative entropy, we obtain
$S\left( \hat{\rho}_{A_1 \cdots A_N} \Vert \Phi \left( \hat{\rho}_{A_1 \cdots A_N}\right) \right) -
\sum_{j=1}^{N} S\left( \hat{\rho}_{A_j} \Vert \Phi _{j}\left( \hat{\rho}_{A_j}\right) \right) \geqslant 0$.
so, we proved that $\mathcal{D_{\gamma}}\left( \hat{\rho}_{A_1 \cdots A_N}  \right) \geqslant 0$. For other $\gamma$, the proof remain the same, then we get that $\mathcal{D}\left( \hat{\rho}_{A_1 \cdots A_N}  \right) \geqslant 0$.

\end{proof}

{\bf Definition 2.} a  multipartite  state is said to be {\bf genuine multipartite semiquantum (GMS for short)}, if there exists a partition $\gamma$, such that $\hat{\rho}_{1 \cdots N} = \Phi^{\gamma}_{1, 2,..., N}\left( \hat{\rho}_{1 \cdots N} \right)$, where {\bf the super operator $ \Phi^{\gamma}_{1, 2,..., N}$ is defined as Eq.~(\ref{super})},
which means that semiquantum  states are not disturbed by a suitable local measurements, (see Ref \cite{Luo:08-2}for bipartite case).
Indeed, this definition of a semiquantum state implies that $\hat{\rho}_{\gamma}=\Phi _{\gamma} \left( \hat{\rho}_{\gamma}\right)$, which means $\mathcal{D}\left( \hat{\rho}_{1 \cdots N} \right)=0$.

{\bf Remark.}Note that, the genuine multipartite semiquantum  state may contain bipartite quantum discord. To see this, Define $\rho_{ABC}=\rho_{AB}\otimes\rho_{C}$,with $\rho_{AB}$ be the maximal entanglement Bell state, and $\rho_{C}=|\Phi\rangle\langle\Phi|$, $|\Phi\rangle=(1,0)^{T}$. The 3-partite state $\rho_{ABC}$ is a genuine multipartite semiquantum state, but it contain 2-partite quantum discord.

Similar to the proof of Ref \cite{Luo:08-2}, we get the following conditions for a  N partite  state $\rho_{1, 2,..., N}$ to be genuine multipartite semiquantum, the following are equivalent:

(1). $\rho_{1, 2,..., N}$ is genuine multipartite semiquantum.

(2). there exists a partition $\gamma|\gamma^{\prime}$(e.g. for three partite state, $A|BC$ is such a partition), such that the state do not changed after the action of the superoperator  $\Phi^{\gamma}_{1, 2,..., N}$, which is given by
$\Phi^{\gamma}_{1, 2,..., N}\left( \hat{\rho}_{1, 2,..., N}\right) =\sum_{k} \left(I_{\gamma^{\prime}}\otimes
\hat{\Pi}_{\gamma}^{k} \right) \hat{\rho}_{1, 2,..., N} \left(I_{\gamma^{\prime}}\otimes \hat{\Pi}_{\gamma}^{k}\right) \, .$.

(3). $\rho_{1, 2,..., N}$ commutes with each  $I_{\gamma^{\prime}}\otimes \hat{\Pi}_{\gamma}^{k}$.

(4).$\rho_{1, 2,..., N}$ has the representation as: $$\rho_{1, 2,..., N}=\sum\limits_{k} p_{k}\rho_{\gamma^{\prime}}^{k}\otimes \hat{\Pi}_{\gamma}^{k}$$, with $p_{k}$ a probability distribution, and  $\rho_{\gamma^{\prime}}^{k}$ are local states on subsystem  $\gamma^{\prime}$, e.g., $\gamma^{\prime}=23$.

In \cite{Yu}, the authors get a witness to detect the semi-quantum  of bipartite state, as follows:

For a given bipartite state $\varrho_{AB}$, a polynomial LU invariant of degree $k$  is given by $\mathrm{Tr}(U^AU^B\varrho^{\otimes k}_{AB})$, where $U^{A(B)}$ is some permutation operator acting on $k$ copies of subsystem $A(B)$ \cite{invariants}. In what follows we shall consider only $k=4$ copies of the state and label them with numbers from 1 to 4. As examples the permutation operator $U^B$ may be taken as $V^B_{12} V_{34}^B$ or $V^B_{13} V_{24}^B$,  where
\begin{equation}
V_{ij}^B=\sum_{n_1,n_2=0}^{d_B-1}|n_1,n_2\rangle\langle n_2,n_1|_{ij}=\sum_{\mu=0}^{d_B^2-1}G_\mu^{B_i}\otimes G_\mu^{B_j}
\end{equation}
is the swapping operator acting on two copies of qudit $B$ labeled with  $i,j=1,2,3,4$. Here we have introduced  a complete set of local orthogonal observables \cite{LOO} $\{G_\mu^{B}\mid \mu=0,1,2,\ldots d^2_{B}-1\}$ satisfying $\mathrm{Tr}(G_\mu^{B} G_\nu^{B})=\delta_{\mu\nu}$ for qudit $B$. The permutation operator $U^A$ may be the cyclic permutation operator
\begin{equation}
X_A=\sum_{n_1,n_2,n_3,n_4=0}^{d_A-1}|n_1,n_2,n_3,n_4\rangle\langle n_2,n_3,n_4,n_1|
\end{equation}
acting on four copies of qudit $A$. It is obvious that $X_A=V_{12}^AV_{23}^AV_{34}^A$. Our main result is:

{\it Lemma \cite{Yu} } A $d_A\times d_B$ bipartite state $\varrho_{AB}$ has a vanishing quantum discord, i.e, $D_A(\varrho_{AB})=0$, if and only if $\tr (W\varrho_{AB}^{\otimes 4})=0$ where
\begin{equation}
W=\frac12(X_A+X_A^\dagger)(V_{13}^BV_{24}^B-V_{12}^BV_{34}^B).
\end{equation}

Now, we turn to our problem, we wish to get a witness to detect the semi-quantum  of N partite state.

To do this, if we split the  N partite state as a two partite state, that is, $\gamma|\gamma^{\prime}$(e.g. for three partite state, $A|BC$ is such a partition), look $\gamma$ as part A, $\gamma^{\prime}$ as part B, then we can get the following:

{\it Theorem 2 }Regard  an  N partite  state $\rho_{1, 2,..., N}$ as  a bipartite state $\varrho_{A^{\gamma}B^{\gamma^{\prime}}}$, (with that look $\gamma$ as part A, $\gamma^{\prime}$ as part B, so A and B only depend on $\gamma,\gamma^{\prime}$, and we use $A^{\gamma},B^{\gamma^{\prime}}$ to denote this ), then $\rho_{1, 2,..., N}$ has a vanishing quantum discord, i.e, $D(\rho)=0$, if and only if there exists a partition $\gamma|\gamma^{\prime}$, such that  $\tr (W\varrho_{A^{\gamma}B^{\gamma^{\prime}}}^{\otimes 4})=0$ where
\begin{equation}
W=\frac12(X_{A^{\gamma}}+X_{A^{\gamma}}^\dagger)(V_{13}^{B^{\gamma^{\prime}}}V_{24}^{B^{\gamma^{\prime}}}-
V_{12}^{B^{\gamma^{\prime}}}V_{34}^{B^{\gamma^{\prime}}}).
\end{equation}

So the witness can be find as: $\min\limits_{\gamma}\tr (W\varrho_{A^{\gamma}B^{\gamma^{\prime}}}^{\otimes 4})$, where the minimal run over all possible partition $\gamma$.

%%%%%%%%%%%%%%%%%%%%%%%%%%%%%

\end{document}